\documentclass[conference]{IEEEtran}
\IEEEoverridecommandlockouts
\usepackage{cite}
\usepackage{amsmath,amssymb,amsfonts}
\usepackage{algorithmic}
\usepackage{booktabs}
\usepackage{multirow}
\usepackage{makecell}
\usepackage{graphicx}
\usepackage{textcomp}
\usepackage{xcolor,colortbl}

\usepackage{adjustbox}
\usepackage{threeparttable}

\usepackage{hyperref}
\usepackage[skins]{tcolorbox}

\usepackage{xspace}
\xspaceaddexceptions{],)}
\newcommand{\paul}{S1\xspace}
\newcommand{\sathish}{S2\xspace}
\newcommand{\tao}{S3\xspace}
\newcommand{\geens}{S4\xspace}
\newcommand{\gonzalez}{S5\xspace}
\newcommand{\alizadeh}{S6\xspace}

\newcommand{\accuracy}{C1\xspace}
\newcommand{\fscore}{C2\xspace}
\newcommand{\storageSize}{RE1\xspace}
\newcommand{\gpuUtil}{RE2\xspace}
\newcommand{\gpuMem}{RE3\xspace}
\newcommand{\gpuPower}{RE4\xspace}
\newcommand{\gpuEnergy}{RE5\xspace}
\newcommand{\inferenceLatency}{RE6\xspace}
\newcommand{\inferencePower}{RE7\xspace}
\newcommand{\inferenceEnergy}{RE8\xspace}

\def\BibTeX{{\rm B\kern-.05em{\sc i\kern-.025em b}\kern-.08em
    T\kern-.1667em\lower.7ex\hbox{E}\kern-.125emX}}

\begin{document}

\title{Aggregating empirical evidence from data strategy studies: a case on model quantization\\
\thanks{This work is partially supported by the GAISSA project TED2021-130923B-I00, funded by MCIN/AEI/10.13039/501100011033 and the European Union Next Generation EU/PRTR. It is also partially funded by the Joan Or{\'o} pre-doctoral support program (BDNS 657443), co-funded by the European Union. Prof. Travassos is a Brazilian CNPq researcher and CNE Faperj.}
}

\author{\IEEEauthorblockN{Santiago {del Rey}\IEEEauthorrefmark{1}, {Paulo S{\'e}rgio Medeiros} {dos Santos}\IEEEauthorrefmark{2}, {Guilherme Horta} Travassos\IEEEauthorrefmark{3},\\Xavier Franch\IEEEauthorrefmark{1} and Silverio Mart{\'i}nez-Fern{\'a}ndez\IEEEauthorrefmark{1}}
\IEEEauthorblockA{
    \IEEEauthorrefmark{1}\textit{Universitat Polit{\`e}cnica de Catalunya (BarcelonaTech)}, Barcelona, Spain\\
    Email:\{santiago.del.rey, xavier.franch, silverio.martinez\}@upc.edu
}
\IEEEauthorblockA{
\IEEEauthorrefmark{2}\textit{UNIRIO}, Rio de Janeiro, Brazil \\
Email: pasemes@uniriotec.br}
\IEEEauthorblockA{\IEEEauthorrefmark{3}\textit{PESC/COPPE/UFRJ}, Rio de Janeiro, Brazil \\
Email: ght@cos.ufrj.br}
}


\maketitle

\begin{abstract}
\underline{Background}: As empirical software engineering evolves, more studies adopt data strategies---approaches that investigate digital artifacts such as models, source code, or system logs rather than relying on human subjects. Synthesizing results from such studies introduces new methodological challenges. \underline{Aims}: This study assesses the effects of model quantization on correctness and resource efficiency in deep learning (DL) systems. Additionally, it explores the methodological implications of aggregating evidence from empirical studies that adopt data strategies. \underline{Method}: We conducted a research synthesis of six primary studies that empirically evaluate model quantization. We applied the Structured Synthesis Method (SSM) to aggregate the findings, which combines qualitative and quantitative evidence through diagrammatic modeling. A total of 19 evidence models were extracted and aggregated. \underline{Results}: The aggregated evidence indicates that model quantization weakly negatively affects correctness metrics while consistently improving resource efficiency metrics, including storage size, inference latency, and GPU energy consumption---a manageable trade-off for many DL deployment contexts. Evidence across quantization techniques remains fragmented, underscoring the need for more focused empirical studies per technique. \underline{Conclusions}: Model quantization offers substantial efficiency benefits with minor trade-offs in correctness, making it a suitable optimization strategy for resource-constrained environments. This study also demonstrates the feasibility of using SSM to synthesize findings from data strategy-based research.
\end{abstract}


\section{Introduction}

Software engineering (SE) increasingly relies on data strategy studies~\cite{storeyWhoWhatHow2020} to understand and improve software development and deployment practices. Data strategies refer to ``empirical studies that rely primarily on archival, generated or simulated data''~\cite{storeyWhoWhatHow2020}, using a wide range of specific methods, including experiments and data mining studies. The rise of data-intensive paradigms and technologies such as the Internet of Things (IoT) and Artificial Intelligence (AI) has led to empirical research using data strategies to evaluate tools, predict system behaviors, and optimize software processes, referring to a particular phenomenon. Although these studies provide valuable information, they remain largely disconnected, with findings often limited to specific contexts and lacking broader theoretical integration. Therefore, the SE field struggles with few theories and needs more structured syntheses of existing research to guide future advancements. This situation raises a critical question: Can we use this growing body of knowledge to form new theories?

The intersection of SE and AI is an emerging research area where a notable portion of publications use data strategies~\cite{martinez-fernandezSoftwareEngineeringAIBased2022}, making it a suitable field for aggregation studies. Rapid adoption of deep learning (DL) systems has led to many empirical studies on model efficiency, optimization techniques, and deployment strategies. However, these studies often differ in their research methods, datasets, and evaluation criteria, making it difficult to draw generalizable conclusions. Model quantization, a technique to improve resource efficiency in DL systems with little effect on correctness, exemplifies this challenge. Although widely studied, empirical findings on its benefits for resource efficiency remain fragmented and context-dependent. For instance, Paul et al. study the effects of quantizing a DL model from 64-bit floating-point (FP64) to different fixed-point precisions (e.g., 2-bit fixed-point (Q0.2)) in the context of a respiratory anomaly detection system~\cite{paulEnergyEfficientRespiratoryAnomaly2022}. Instead, Gonzalez et al. study how quantizing from FP32 to 8-bit integer (INT8) affects multiple image classification models~\cite{gonzalezImpactMLOptimization2025}. This paper addresses this issue by conducting a research synthesis of empirical studies on model quantization to assess its effects on system correctness and resource efficiency. In addition, we examine the methodological challenges in synthesizing software engineering research that relies on data strategies. To ensure methodological rigor, we use the Structured Synthesis Method (SSM), a systematic approach that can aggregate qualitative and quantitative evidence through diagrammatic models~\cite{medeirosdossantosRepresentationAggregationEvidence2013}.

Our contributions are twofold:
\begin{itemize}
    \item We systematically collect and synthesize empirical primary studies on model quantization, identified by J{\"a}rvenp{\"a}{\"a} et al. as a key sustainable SE tactic~\cite{jarvenpaaSynthesisGreenArchitectural2024}, analyzing its impact on DL systems' correctness and resource efficiency. By means of aggregation, we propose a theory of the role of model quantization in optimizing DL systems.
    \item We share lessons learned and methodological challenges of aggregating empirical studies based on data strategies that can inform future empirical syntheses.
\end{itemize}

\textbf{Data availability}. A complete replication package is available on~\cite{ReplicationPackage}. The evidence and aggregated models are publicly accessible in the Evidence Factory tool~\cite{EvidenceFactorySynthesis}.

\section{Background}
\subsection{Model quantization}
Model quantization is one of the most commonly used techniques for compressing and accelerating DL models, particularly in industrial applications, due to its stable compression ratio and performance gains~\cite{liContemporaryAdvancesNeural2024}. At its core, quantization maps values from a continuous, infinite domain into a smaller set of discrete finite values~\cite{rokhComprehensiveSurveyModel2023}. This transformation significantly reduces memory consumption and accelerates inference, making it crucial for deploying DL models in resource-constrained environments such as IoT devices and embedded systems.

Model quantization involves three key considerations: (i) target precision level (e.g., INT8); (ii) quantized components, which are typically the model's weights, activations, or both~\cite{rokhComprehensiveSurveyModel2023}; and (iii) quantization timing, which follows one of two primary approaches: Quantization-Aware Training (QAT) where quantization is integrated into the training process, allowing the network to learn with quantized values from the start; and Post-Training Quantization (PTQ) where quantization is applied after training on full-precision data.

Recent studies have surveyed advancements in model quantization, consistently reporting reductions in model size, memory footprint, and computational demand. These benefits enable efficient deployment on edge devices~\cite{rokhComprehensiveSurveyModel2023, liContemporaryAdvancesNeural2024}. While quantization often introduces a trade-off in model accuracy, QAT can maintain performance close to full-precision baselines and is effective even in sensitive domains such as medical imaging, albeit requiring retraining~\cite{paddoScopingReviewQuantization2024}. PTQ, on the other hand, offers greater deployment efficiency but may result in more significant accuracy degradation, particularly at lower bit-widths or with novel model architectures~\cite{rokhComprehensiveSurveyModel2023, liContemporaryAdvancesNeural2024}. Advanced techniques such as mixed-precision, non-uniform schemes, and data-free quantization are being actively explored to optimize the efficiency-accuracy balance under various constraints~\cite{rokhComprehensiveSurveyModel2023, liContemporaryAdvancesNeural2024}. However, the findings across existing studies vary significantly depending not only on the core quantization parameters discussed above, but also on factors such as the DL architecture used, datasets involved, and the target hardware platform. As such, generalizing results beyond the original experimental settings remains challenging.

To the best of our knowledge, this is the first study to systematically synthesize empirical evidence on the effects of model quantization, providing a structured understanding of its impact on resource efficiency and correctness.

\subsection{Aggregating evidence from data strategy studies}
Research synthesis is a cornerstone of knowledge creation in empirical SE 
as it aims to summarize, integrate, and interpret findings from multiple primary studies on specific topics~\cite{santosResearchSynthesisSoftware2020}.

However, the landscape of empirical research in SE is diverse. Storey et al. proposed a framework categorizing empirical studies based on their primary beneficiary (Who), contribution type (What), and research strategy (How). Their framework highlights a significant portion of SE research employing data strategies---studies relying primarily on archival, generated, or simulated data (e.g., from software repositories, system logs, historical datasets) often without direct, concurrent involvement of human participants. While these data strategy studies are prevalent, particularly in venues like ICSE and EMSE~\cite{storeyWhoWhatHow2020}, the methods for synthesizing their specific types of evidence seem less established than those for human-centric studies like experiments or surveys.

Existing synthesis work published in ESEM and EMSE showcases various approaches. Meta-analysis is commonly used to aggregate quantitative results, often from controlled experiments, as demonstrated by Scanniello et al.~\cite{scannielloSoftwareModelsBased2018} in their aggregation of 12 experiments on UML and source code comprehensibility, where they explicitly addressed study heterogeneity.  Abou Khalil and Zacchiroli~\cite{aboukhalilSoftwareArtifactMining2022} performed a meta-analysis focused on what artifacts are mined in SE conferences (many using data strategies), but not on how to synthesize the results derived from mining these artifacts. Specific synthesis methodologies like the Structured Synthesis Method (SSM) have been proposed~\cite{medeirosdossantosRepresentationAggregationEvidence2013} and applied to combine both quantitative and qualitative evidence, for instance, on software development productivity~\cite{chapettaEvidencebasedTheoreticalFramework2020}, often using SLRs as a starting point. Syntheses have also focused on human-centric topics, such as the well-being of software engineers, primarily aggregating results from quantitative survey studies~\cite{godliauskasWellbeingSoftwareEngineers2024}. Furthermore, meta-research efforts have compared different techniques for aggregating families of interrelated replications~\cite{santosComparingTechniquesAggregating2018} or analyzed the characteristics of SLRs themselves~\cite{wangHowManyPapers2023}. Ribeiro et al.\cite{ribeiroInvestigationEmpiricalContradictions2023} contributed to the discussion of empirical aggregation by investigating contradictions across local readability studies, and identifying sources of variation in study design and measurement approaches.

Despite the aforementioned relevant synthesis activities, there appears to be a gap in research synthesis specifically addressing the aggregation of evidence from data strategy studies. The peculiarities of these studies (e.g., non-human subjects, increased number of independent variables) may require adapted synthesis guidelines or techniques. Standard meta-analysis focusing on effect sizes from human experiments or surveys might not always be directly applicable or sufficient for synthesizing the diverse outputs of data strategy research (e.g., predictive model performance, descriptive statistics from repositories, correlation findings). While methods like SSM~\cite{medeirosdossantosRepresentationAggregationEvidence2013} offer flexibility, dedicated guidelines for synthesizing evidence predominantly derived from data strategies, considering their unique validity threats and result types, are currently lacking. This highlights the need for methodological research to develop and validate synthesis approaches tailored to the nuances of data-driven empirical software engineering research.

\section{Methodology}
We aim to contribute to the topic of resource efficiency of DL systems~\cite{ResourceefficientSoftwareFocus}, which is framed into the broader area of environmental sustainability. Our goal is defined as:

\textbf{Analyze} \textit{the use of model quantization}
\textbf{for the purpose of} \textit{characterizing its effects}
\textbf{with respect to} \textit{correctness (Accuracy, and F1 Score) and resource efficiency (Storage size, GPU utilization, GPU memory utilization, GPU power draw, GPU energy consumption, Inference latency, Inference power draw, and Inference energy consumption)}
\textbf{from the point of view of} \textit{DL engineers}
\textbf{in the context of} \textit{aggregating primary studies evaluating model quantization in DL systems}

\begin{figure}[t]
    \centering
    \includegraphics[width=0.7\columnwidth]{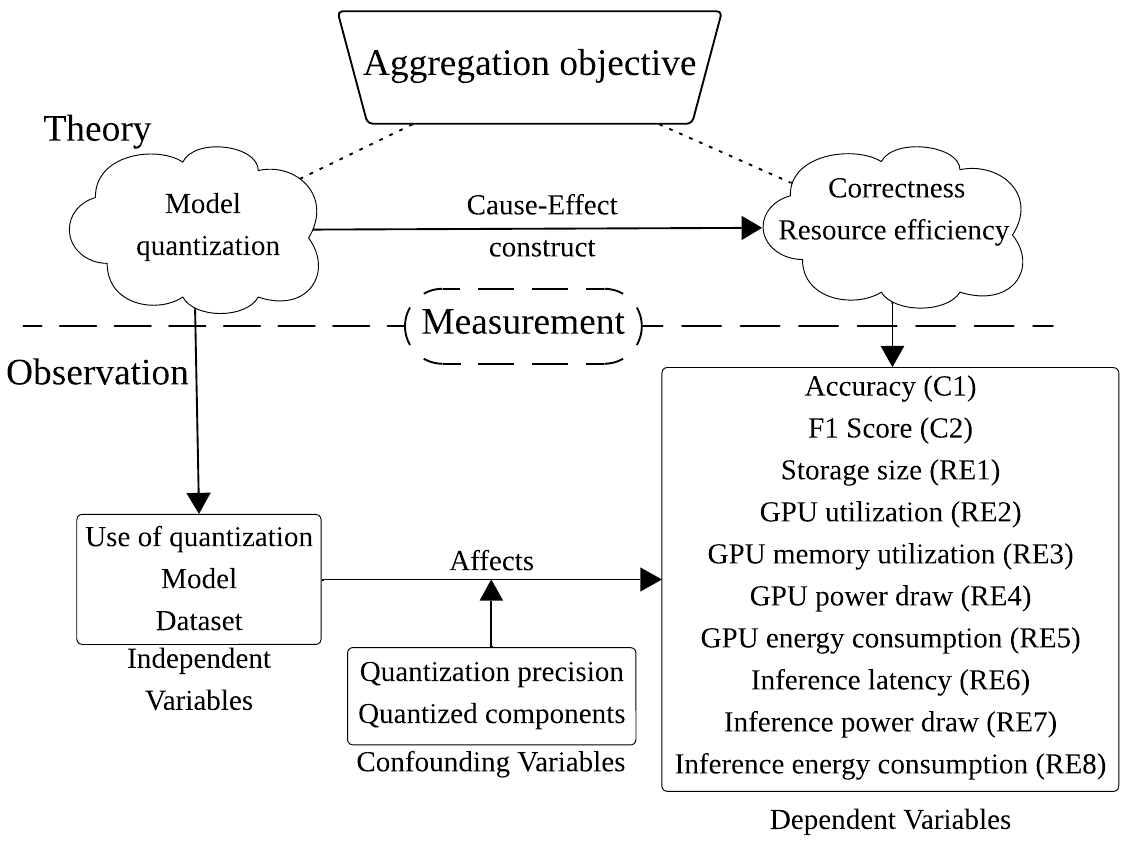}
    \caption{Experiment principles of the study (adapted from Wohlin et al.~\cite{wohlinExperimentationSoftwareEngineering2012}).}\label{fig:experiment-principles}
\end{figure}

From this goal, we derive the following research question (RQ): \emph{``How does model quantization affect system correctness and resource efficiency when deploying DL systems?''}. 
Fig.~\ref{fig:experiment-principles} shows the basic experiment principles behind this study.

\begin{table*}
\centering
\begin{minipage}{\textwidth}
\centering
\caption{Selected primary studies.}\label{tab:studies}
\begin{adjustbox}{width=\textwidth}
\rowcolors{2}{}{gray!25}
\begin{tabular}{@{\kern\tabcolsep}lp{2.2cm}llllp{1.8cm}lp{2.7cm}llll@{\kern\tabcolsep}}
\toprule
\rowcolor{gray!50}
 &  &  &  & & \multicolumn{2}{c}{Quantization Precision} & & & & & &\\
\rowcolor{gray!50}
\multicolumn{1}{c}{\multirow{-2}{*}{Study Id.}} &
  \multicolumn{1}{c}{\multirow{-2}{*}{\makecell[cc]{Study Type:\\Instruments}}} &
  \multicolumn{1}{c}{\multirow{-2}{*}{\makecell[cc]{Application\\Domain}}} &
  \multicolumn{1}{c}{\multirow{-2}{*}{\#Models}} &
  \multicolumn{1}{c}{\multirow{-2}{*}{\#Datasets}} &
  \multicolumn{1}{c}{Baseline} & 
  \multicolumn{1}{c}{Target\footnote{W refers to apply the target quantization precisions only to weights, A for only activations, and F for both weights and activations.}} &
  \multicolumn{1}{c}{\multirow{-2}{*}{\makecell[cc]{Quantization\\Timing}}} &
  \multicolumn{1}{c}{\multirow{-2}{*}{\makecell[cc]{Studied\\Effects}}} &
  \multicolumn{1}{c}{\multirow{-2}{*}{Data quality}} &
  \multicolumn{1}{c}{\multirow{-2}{*}{\makecell[cc]{Primary\\Belief}}}&
  \multicolumn{1}{c}{\multirow{-2}{*}{\makecell[cc]{\#Theoretical\\Structures}}}&
  \multicolumn{1}{c}{\multirow{-2}{*}{Year}} \\ \midrule
\paul \cite{paulEnergyEfficientRespiratoryAnomaly2022} & \makecell[lc]{Experiment\\(observational):\\numerical estimation} & \makecell[lc]{Respiratory anomaly\\detection} & 1 & 1 & FP64 &        \makecell[lc]{Q0.2, Q0.4, Q0.8,\\Q0.16, Q0.32\\(F)}   & QAT &  \accuracy, \storageSize, \inferenceEnergy &   Comparative & 37\% & 5 & 2022 \\
\sathish \cite{sathishVerifiableEnergyEfficient2022} & \makecell[lc]{Experiment\\(observational):\\numerical estimation} &\makecell[lc]{Thorax disease classification,\\Liver disease segmentation} & 4 & 2 &          FP32 &        INT8 (F)   & PTQ & \accuracy, \storageSize, \inferenceEnergy &   Comparative & 39\% & 1 & 2022 \\
\tao \cite{taoExperimentalEnergyConsumption2022} & \makecell[lc]{Experiment\\(quasi-experiment):\\hardware-based} & Bird call classification & 1 & 1 &          FP32 & \makecell[lc]{Q0.8, Q0.16\\(W, A, F)}   & QAT & \accuracy, \fscore, \storageSize, \inferenceLatency, \inferencePower, \inferenceEnergy &   Comparative & 64\% & 8 & 2022 \\
\geens \cite{geensEnergyCostModelling2024} & Non-systematic: numerical estimation & N/A & 1 & N/A & FP32& INT1, INT4 (W)   & N/A & \inferenceLatency, \inferenceEnergy &   \makecell[lc]{Comparative\\ (Bar chart)}& 18\% & 2 & 2024 \\
\gonzalez \cite{gonzalezImpactMLOptimization2025} & \makecell[lc]{Experiment\\(quasi-experiment):\\hardware-based,\\nvidia-smi} & Image classification & 28 & 5 & FP32 &        INT8 (F)  & PTQ & \accuracy, \storageSize, \gpuUtil, \gpuPower, \gpuEnergy \inferenceLatency, \inferencePower, \inferenceEnergy &   Precise& 74\% & 1 & 2025 \\
\alizadeh \cite{alizadehLanguageModelsSoftware2025} & \makecell[lc]{Experiment\\(quasi-experiment):\\pyNVML, pyRAPL} & Code generation & 18 & 1 &          FP16&        INT4, INT8 (W)  & PTQ  &  \accuracy, \storageSize, \gpuUtil, \gpuMem, \gpuPower, \gpuEnergy \inferenceLatency, \inferencePower, \inferenceEnergy & Precise & 71\% & 2 & 2025 \\ \bottomrule
\end{tabular}%
\end{adjustbox}
\end{minipage}
\end{table*}

To answer the RQ, we aggregate the results of primary studies on model quantization using SSM. As a synthesis approach that supports both qualitative and quantitative evidence, SSM captures key contextual aspects while indicating the overall direction of effects (e.g., positive or negative) and estimating the associated certainty. This integrative and interpretive nature of SSM was the main reason for selecting it as our synthesis method, given the variability in experimental designs and variables among model quantization studies. SSM has been applied across various research strategies, including respondent, field, and lab strategy studies~\cite{martinez-fernandezAggregatingEmpiricalEvidence2015, dossantosBenefitsChallengesUsing2018}.

After defining the goal, an SSM study has three main steps~\cite{medeirosdossantosRepresentationAggregationEvidence2013}. First, primary studies are identified and selected according to predefined criteria. Second, the studies are analyzed, and evidence is extracted and modeled in a diagrammatic representation. Third, the evidence is synthesized. This synthesis reflects the main trends and beliefs of the analyzed evidence. Next, we detail how we applied each step to synthesize the research on model quantization for deploying DL systems.

\subsection{SSM step 1: Selecting primary studies}
We defined a search strategy and inclusion and exclusion criteria to select primary studies.

\subsubsection{Search strategy}
We adapted the search strategy from Wohlin et al.~\cite{wohlinSuccessfulCombinationDatabase2022}, using Scopus for peer-reviewed studies, while omitting the snowballing step. Additionally, we follow the recommendation in \cite{martinez-fernandezSoftwareEngineeringAIBased2022} and include arXiv to capture recent, yet-to-be-reviewed work. To manage the high volume of DL research, we focused on peer-reviewed publications from 2020 onward and preprints from 2022 onward, assuming older preprints likely have a peer-reviewed counterpart.

To search, we used the following search string, which aims to find primary studies of model quantization focusing on the energy consumption of applying this technique:

\emph{(``machine learning'' OR ``ML'' OR ``deep learning'' OR ``DL'' OR ``large language model'' OR ``LLM?'' OR ``neural network'' OR ``?NN'' OR ``fundational model'' OR ``agent'') AND (``quantization'' OR ``quantize'' OR ``quantized'') AND (``energy consumption'' OR ``energy efficien*'' OR ``sustain*'' OR ``carbon footprint'' OR ``carbon emission'') AND NOT (``FL'' OR ``federated learning'')}

The search used filters on the titles and abstracts of the studies. We excluded papers on federated learning since this technique involves training models, which violates the third inclusion criterion in the next section.

\subsubsection{Inclusion and exclusion criteria}
The inclusion of a study is considered using five criteria: (i) it regards the use of model quantization to optimize a DL model, (ii) it regards the environmental sustainability and/or energy efficiency of using model quantization, (iii) it analyzes the use of model quantization for model inference, (iv) it regards the use of model quantization at the software level, and (v) it controls the factors in each trial avoiding free variation among runs.

Regarding the exclusion, we define five exclusion criteria: (i) it combines model quantization with other optimization techniques, (ii) it does not report a non-quantized baseline, (iii) it is secondary or tertiary, (iv) it is not written in English, and (v) it is in the form of editorials, tutorials, books, extended abstracts, or any grey literature. 

\subsubsection{Study selection}

The search performed in February 2025 retrieved 936 studies. We used the Gemini 2.0 Flash Large Language Model (LLM) to speed up the filtering process and narrow the list of retrieved studies. We followed the strategy reported in~\cite{felizardoChatGPTApplicationSystematic2024} and complemented it with the suggestions of the PRIMES framework~\cite{demartinoFrameworkUsingLLMs2024}. After the LLM filtering, 245 studies remained. We refer to the replication package for more details on how the LLM filtering was done.

From the studies suggested by the LLM, we performed a subsequent manual assessment based on their titles and abstracts, reducing the number of studies to 86. Six were chosen for data extraction after a full reading of the primary studies. The most important details of the selected studies are presented in Table~\ref{tab:studies}.

\subsection{SSM step 2: Analysis and evidence representation}\label{sec:evidence-extraction}

This step aims to bring the selected primary studies into the same perspective so they can be aggregated. In SSM, studies are represented by evidence models, which are considered their \textit{theoretical structure} and describe the contextual aspects and the effects expected from the object of study---model quantization in our case.

\begin{figure*}[tb]
    \centering
    \includegraphics[width=0.85\textwidth]{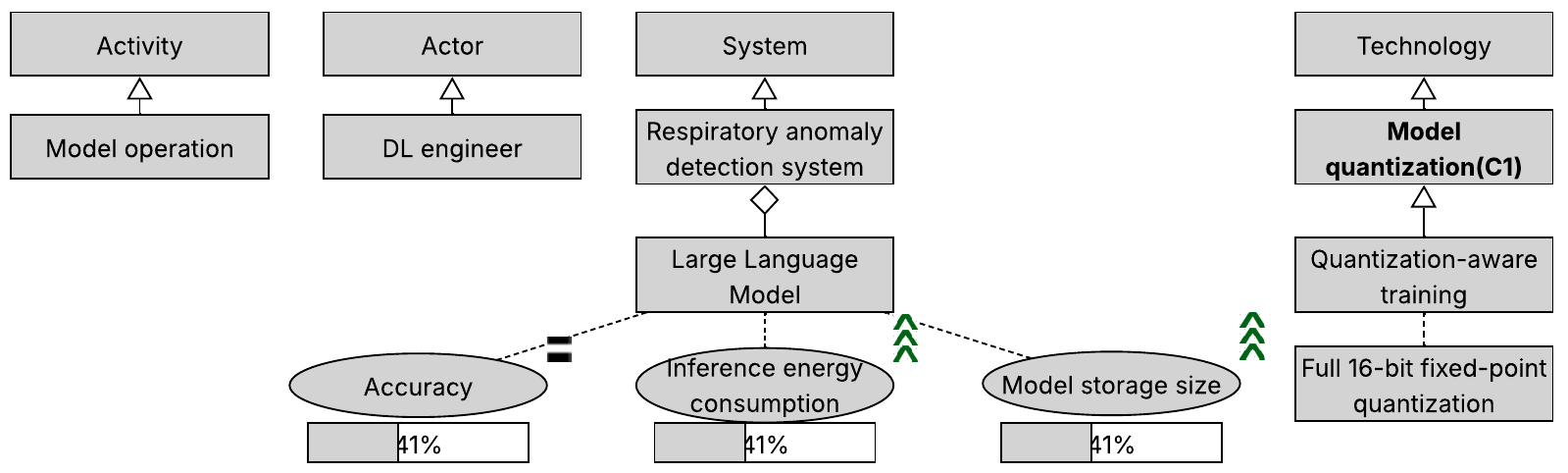}
    \caption{Evidence model representing the results of \href{https://evidencefactory.lens-ese.cos.ufrj.br/evidenceEditor/263147}{\paul}.}
    \label{fig:example-evidence}
\end{figure*}

The selected studies reported evidence of quantizing different components to multiple precision levels. After extracting and analyzing the data in these studies (see details below), we created 19 \textit{theoretical structures} accounting for all the combinations of precision and quantized components present in the studies. Furthermore, we elicited 36 semantic constructs used in these theoretical structures. Fig.~\ref{fig:example-evidence} shows a subset of them. The representation has three possible types of ``structural'' relationships: \textit{is a},  \textit{part of}, and \textit{property of}. 
The \textit{is a} and \textit{part of} relationships use the same UML notation for generalization and composition. Dashed connections denote \textit{properties}. The relationships link two types of concepts -- \textit{value} and \textit{variable}.

A \textit{value} concept represents a specific value of a variable, typically an independent variable, and is visually depicted as a rectangle in conceptual diagrams. These concepts are classified into \textit{archetypes} (the root of each hierarchy), \textit{causes} (denoted with bold font and a ``C1'' following the name (e.g., ``Model quantization''), and \textit{contextual aspects} (e.g., ``Respiratory anomaly detection system''). 

A \textit{variable} concept represents variations in value, typically associated with a dependent variable. These concepts are depicted using ellipses for \textit{effects} (e.g., ``Inference energy consumption''). 
As the conceptual framework implies the existence of effects, these are not explicitly connected to their causes through lines in the diagram. \textit{Cause-effect} relationships are classified as \textit{influence} relationships.

A seven-point Likert scale is used to express effect size, ranging from strongly negative to strongly positive, and is displayed above the ellipse representing the effect. The number of arrows indicates the strength of the effect, \includegraphics[height=1.5ex]{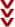} indicates strongly negative, and \includegraphics[height=1.5ex]{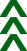} strongly positive. Half arrows indicate a range (e.g., \includegraphics[height=1.5ex]{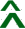} indicates an effect between weakly positive and positive), and an \includegraphics[height=1.5ex]{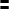} indicates an indifferent effect. For example, in Fig.~\ref{fig:example-evidence} \includegraphics[height=1.5ex]{figures/effect-intensities/SP.png} indicates that ``Model storage size'' is strongly positively affected by ``Model quantization'', while \includegraphics[height=1.5ex]{figures/effect-intensities/I.png} indicates ``Accuracy'' is indifferently affected by ``Model quantization''. The last aspect regarding \textit{variable} concepts is the \textit{belief value}, ranging from 0\% to 100\% (or 0 to 1), to quantify confidence in the observed \textit{effects}. \textit{Belief values} are visually represented as bars placed under each element, e.g., ``Accuracy'' may have a \textit{belief value} of 41\%, indicating the degree of confidence in the measured effect.

SSM provides the Evidence Factory tool to facilitate the modeling and synthesis of evidence~\cite{dossantosStructuredSynthesisMethod2017}. This instrument offers online diagrammatic capabilities for modeling the evidence derived from the selected studies. The terminology employed to describe evidence is maintained within a global glossary, accessible to all further evidence generated. In the case of cause-effect and moderation relationships, the associated uncertainty is evaluated based on the type of study and its quality assessment. The tool incorporates a questionnaire for this evaluation, which is required to be completed for each study. Lastly, it aids in integrating evidence models and produces an aggregated evidence model.

\begin{figure*}[p]
    \centering
    \includegraphics[height=\textheight]{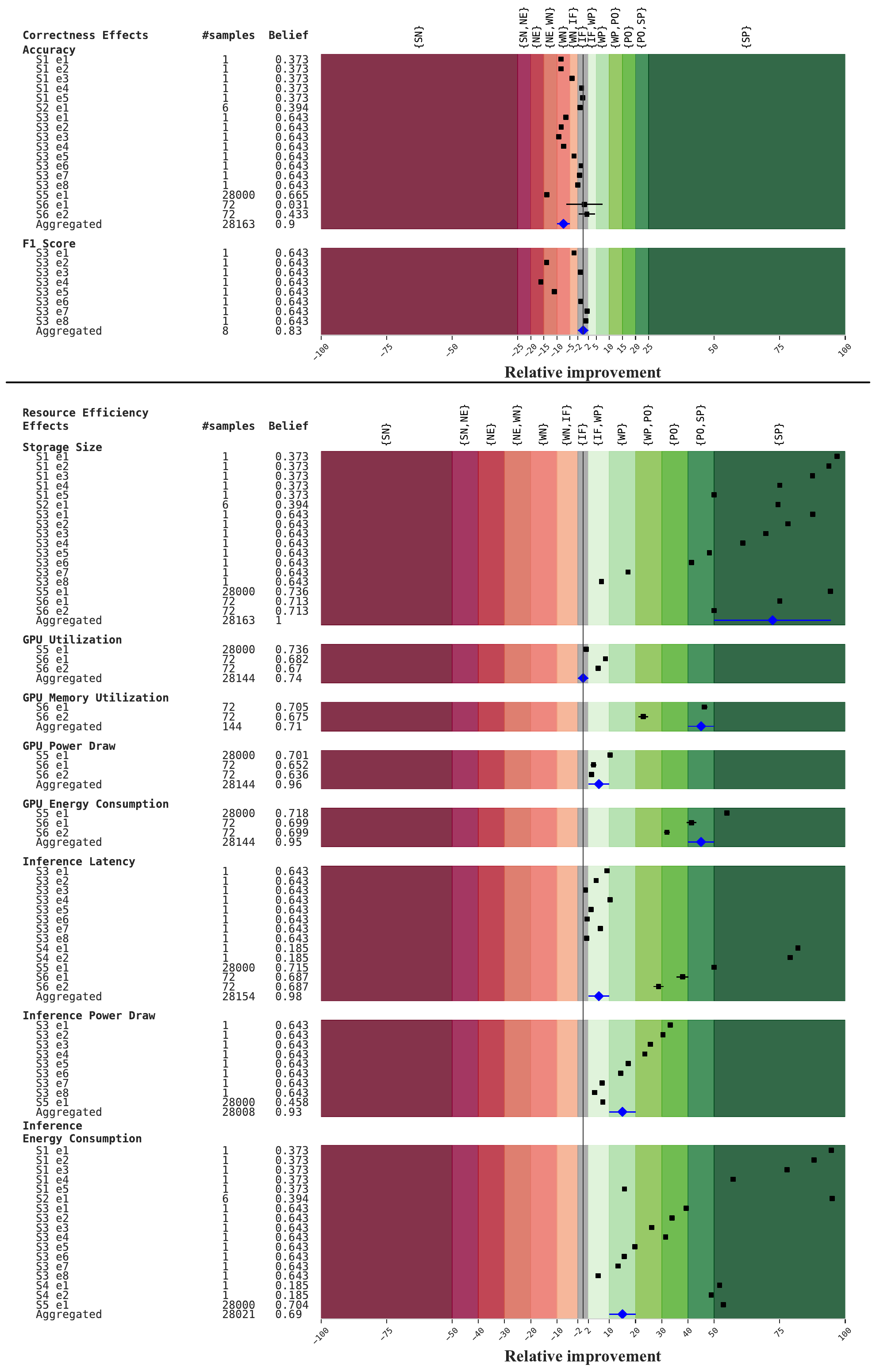}
    \caption{Adapted forest plot showing the \textit{effects} of applying model quantization on dependent variables from all the extracted evidence from primary studies S1 to S6. Links to each evidence diagram are provided in the replication package.}\label{fig:forestplot}
\end{figure*}

\textbf{Extracting information to build evidence models}:
The creation of evidence models begins with extracting information from the studies to define key concepts---contextual aspects, and effects---and to estimate \textit{belief values} for the latter. This process unfolds in two stages: first, determining concepts and relationships, and second, estimating confidence. The initial stage is similar to the coding process~\cite{auerbachQualitativeDataIntroduction2003}, but focused on directly mapping elements into the theoretical structure. A key technique in this stage is the translation procedure~\cite{brittenUsingMetaEthnography2002}, which enables evidence aggregation by aligning similar but differently labeled concepts under a common term. For instance, in our context, ``Large Language Model'' may be generalized as ``DL model'' to facilitate comparability. Given the loss of contextual information, such generalization must be carefully evaluated case by case. Besides modeling concepts, effect intensity and direction are determined by systematically converting quantitative data into the Likert framework using contextually defined thresholds (see Fig.~\ref{fig:forestplot}). The second step determines the \textit{effects}' \textit{belief values}. This assessment relies on two primary inputs: the study type and its quality evaluation. The study type is classified using the GRADE evidence hierarchy~\cite{gradeworkinggroupGradingQualityEvidence2004}, which divides the 0–1 belief value range into four sub-ranges: unsystematic observations [0.00–0.25], observational studies [0.25–0.50], quasi-experiments [0.50–0.75], and randomized controlled trials [0.75–1.00]. The quality assessment refines this classification within a 0.25 sub-range, using two checklists outlined in~\cite{medeirosdossantosRepresentationAggregationEvidence2013}.

Table~\ref{tab:studies} shows the study type for the selected papers. We classify the systematic studies using the framework proposed by Storey et al.~\cite{storeyWhoWhatHow2020}. In parenthesis we show the study type in SSM. We classify \paul and \sathish as observational studies given their use of numerical estimations for energy consumption measurements. Details regarding the quality assessment for each study can be found in the Evidence Factory~\cite{EvidenceFactorySynthesis} and the replication package.

To build the evidence models, the first author thoroughly read the papers and extracted the relevant information to create the models. Next, the first and the last two authors reviewed the models, including several meetings to reach a common understanding of the models. All authors performed a final revision of the models and the resulting aggregated model.


\subsection{SSM step 3: Evidence synthesis}\label{sec:evidence-synthesis}
This step aggregates evidence from the evidence models, which requires compatible theoretical structures. In SSM, compatibility occurs when the \textit{value concepts} are the same or have the same meaning, including the \textit{cause}, \textit{archetypes}, and \textit{contextual aspects}. Compatible \textit{theoretical structures} allow their \textit{effects} to be combined based on direction and intensity.

When aggregating multiple evidence models, pairwise comparisons are used to assess their compatibility. If the models do not share identical concepts (e.g., ``DL model''), the researcher must determine whether their meanings are sufficiently similar to justify aggregation. If so, the evidence can still be combined by joining concepts. Alternatively, if the concepts are deemed too distinct, they can be included in the aggregated results, but their effects will not be merged. A third approach applies when a concept appears in only one evidence model, allowing the researcher to decide whether to include or exclude it from the aggregation process.

After determining model compatibility, the aggregation process focuses on pooling \textit{variable concepts} by considering intensity (e.g., positive or negative) and uncertainty (i.e., \textit{belief value}). 
In SSM, the Mathematical Theory of Evidence, known as Dempster-Shafer Theory (DST)~\cite{shafer1976mathematical}, provides the mathematical foundation for combining evidence. DST relies on two primary inputs: the hypotheses believed to have a chance of being true (i.e., those with a \textit{belief value} greater than zero) and the \textit{belief values} themselves. Hypotheses are represented as sets within the powerset of a frame of discernment, which in SSM corresponds to the seven-point Likert scale: $\Theta = \left\{SN, NE, WN, IF, WP, PO, SP\right\}$, where SN stands for ``strongly negative'', IF for ``indifferent'', WP for ``weakly positive'', and so on. Since hypotheses can be singleton sets (e.g., \{PO\}) or compound sets (e.g., \{WN, PO\} indicating uncertainty between ``weakly negative'' and ``positive''), DST structures the combination of evidence, capturing both agreement and imprecision in the aggregated results. Once hypotheses and \textit{belief values} are defined for each evidence model, \textit{Dempster's Rule of Combination} is applied. More details about how DST is used in SSM are available in~\cite{medeirosdossantosRepresentationAggregationEvidence2013}.




\section{Representation of model quantization effects}

Before presenting the aggregated results, we outline the coding process for modeling the theoretical structures and the concepts developed in that process. Given the limited space, we represent only one of the smallest evidence models, extracted from study \paul (see Fig.~\ref{fig:example-evidence}). The study assesses various model quantization techniques to overcome energy limitations of wearable devices for respiratory classification. 

Identifying the \textit{effects} was straightforward as they mapped to the dependent variables, which are clearly reported in the study. Regarding the \textit{effect intensity}, we extracted it directly from the data reported in the study. We compute the average improvement relative to the non-quantized baselines for each effect and then determine the intensity based on the thresholds shown in Fig.~\ref{fig:forestplot}. For instance, the \textit{effect intensity} for ``Inference energy consumption'' was defined as $\{SP\}$ since the average relative improvement was 57.18\%. The \textit{belief values} were based on the study type and the quality assessment described in Sec.~\ref{sec:evidence-extraction}. Additionally, we introduced a DST discount operation to adjust the mass distribution to reflect the source's credibility---a full discount (discount=1) represents a completely unreliable source. In this case, we used Eq.~\ref{eq:discount} to decrease the \textit{belief} of an \textit{effect} based on the ratio between its Interquartile Range (IQR) and its mean. The final \textit{belief} of an \textit{effect} is computed as $base\_belief\times (1-discount)$.

\begin{equation}\label{eq:discount}
    discount=1 - e^{-0.1\times\left|\frac{IQR}{\mu}\right|}
\end{equation}


It is worth noting that several studies reported results of multiple models, datasets, and quantization methods---the combination of quantization precision and quantized components. The \textit{effects} reported in the evidence models are the results of aggregating the metrics by quantization method, since we are interested in the \textit{effects} of ``Model quantization'' independently of the model and dataset. This same procedure applies to all the papers. We aggregate by quantization method and extract multiple \textit{theoretical structures} from each study, due to the studies treating the different methods as independent. We acknowledge that the quantization method is a confounding factor in the aggregation. Ideally, evidence aggregation should be made first for each quantization method and then aggregated under the umbrella of ``Model quantization.'' However, all quantization methods, except one, have a single evidence model, making it impossible to aggregate at this level of abstraction. Hence, we decided to directly perform the aggregation one abstraction level above (i.e., ``Model quantization'').

All the evidence models created for each study can be found in the Evidence Factory tool. To summarize the evidence models, Fig.~\ref{fig:forestplot} shows a forest plot, adapted to SSM, showing the \textit{effects} caused by model quantization for each evidence model. In the first column, we show the \textit{effect} in bold, and below, the evidence reporting the effect. The second column reports the number of samples used to determine \textit{effect} direction and intensity. The third column shows the \textit{belief} associated with the \textit{effect} for that specific evidence. Lastly, the chart displays the mean and the 95\% confidence interval (CI) using squares and horizontal lines, respectively. A vertical line marks 0\% relative improvement. All points to the right of this line indicate a positive improvement, and all points to the left indicate a negative improvement. SSM does not report specific values for aggregated intensities. Hence, the ``Aggregated'' points are located at the center of the resulting intensities, and their CIs cover the respective intensity ranges.

\section{Results}\label{sec:results}

Table~\ref{tab:general-resutls} summarizes the aggregation of the 19 evidence models on model quantization effects emerging from the six primary studies. The columns show: 1) reported effects, 2) primary studies reporting each \textit{effect}, 3) number of studies, 4) number of \textit{evidence models} aggregated, 5) aggregated \textit{effect intensity} indicating \textit{effect} direction (positive/negative), 6) aggregated belief of the \textit{effect}, crucial to the results, 7) conflicts during \textit{effect} aggregation, and 8) differences between the \textit{belief} maximum in individual models and post-aggregation confidence. A positive difference means reinforced effects; a negative one implies conflicting evidence.

\begin{table*}[!t]
\centering
\caption{Aggregated effects of utilizing model quantization using the SSM.}
\label{tab:general-resutls}
\resizebox{\textwidth}{!}{%
\rowcolors{2}{}{gray!25}
\begin{tabular}{@{\kern\tabcolsep}cccccccc@{\kern\tabcolsep}}
\toprule
\rowcolor{gray!50}
 & \multicolumn{7}{c}{Aggregation results} \\ \Xcline{2-8}{1pt}
\rowcolor{gray!50}
\multicolumn{1}{c}{\multirow{-2}{*}{\makecell[cc]{Effect caused by \\ model quantization}}} &
  \multicolumn{1}{c}{\textit{Study Id.}} &
  \multicolumn{1}{c}{\makecell[cc]{\textit{Number}\\\textit{of studies}}} &
  \multicolumn{1}{c}{\makecell[cc]{\textit{\#Evidence}\\\textit{models}}} &
  \multicolumn{1}{c}{\textit{Intensity}} &
  \multicolumn{1}{c}{\textit{Belief}} &
  \multicolumn{1}{c}{\textit{Conflict}} &
  \multicolumn{1}{c}{\textit{Difference}} \\ \midrule
    Accuracy                        & \paul, \sathish, \tao, \gonzalez, \alizadeh           & 5 & 17    & WN        & \textbf{90\%}     & 0.41  & 24\%\\
    F1 score                        & \tao                                                  & 1 & 8     & IF        & 83\%     & 0.62  & 19\%\\ \midrule
    Storage size                    & \paul, \sathish, \tao, \gonzalez, \alizadeh           & 5 & 17    & SP        & \textbf{100\%}    & -     & 26\% \\
    GPU utilization                 & \gonzalez, \alizadeh                                  & 2 & 3     & IF        & 74\%              & -     & 0\%\\
    GPU memory utilization          & \alizadeh                                             & 1 & 2     & \{PO,SP\} & 71\%              & -     & 0\%\\
    GPU power draw                  & \gonzalez, \alizadeh                                  & 2 & 3     & \{IF,WP\} & \textbf{96\%}     & -     & 26\%\\
    GPU energy consumption          & \gonzalez, \alizadeh                                  & 2 & 3     & \{PO,SP\} & \textbf{95\%}     & 0.5  & 23\%\\
    Inference latency               & \tao, \geens, \gonzalez, \alizadeh                    & 4 & 13    & \{IF,WP\} & \textbf{98\%}     & 0.61  & 27\% \\
    Inference power draw            & \tao, \gonzalez                                       & 2 & 9     & WP        & \textbf{93\%}     & 0.64  & 28\% \\
    Inference energy consumption    & \paul, \sathish, \tao, \geens, \gonzalez              & 5 & 17    & WP        & 69\%              & 0.61  & -1.7\% \\
  
 \bottomrule
\end{tabular}%
}
\end{table*}


\subsection{Effects of model quantization that increased their belief}
Seven \textit{effects} caused by model quantization increased their \textit{belief values} after the aggregation. These effects aggregate at least three evidence models. It is important to notice that belief values next to 1 (100\%) are not indicative of certainty, but rather an indication of the evidence strength after aggregation, given the current body of knowledge. Next, we enumerate the results of each of the six \textit{effects}.

Using model quantization weakly negatively affects \textit{accuracy}. Evidence of this \textit{effect} is vast, but there are some contradictory results (see Fig.~\ref{fig:forestplot}). Evidence shows its reduction is minimal and, in specific cases, can even improve.

Model quantization indifferently affects \textit{F1 score}. Although the results come from aggregating eight evidence models, all came from \tao. Similar to \textit{accuracy}, evidence shows contradictory results reporting negative and positive improvements.

Model quantization strongly positively affects \textit{storage size}. Evidence of this \textit{effect} is vast and consensual as reported in Tab.~\ref{tab:general-resutls}. In general, studies reported over 50\% reduction in storage size after applying model quantization (see Fig.~\ref{fig:forestplot}).

Model quantization \{indifferently - weakly positively\} affects \textit{GPU power draw}. Although only three evidence models contain this \textit{effect}, we see a consensus on its direction.

Using model quantization \{positively - strongly positively\} affects \textit{GPU energy consumption} of operating DL models. Although there is some conflict in the aggregated evidence, all agree on the positive effect of model quantization, reporting improvements between 30\% to 55\%.

Model quantization \{indifferently - weakly positively\} affects \textit{inference latency} of DL models. However, we observe a high degree of conflict regarding this effect. While evidence from \tao reports indifferent to weak effects on \textit{inference latency}, evidence from \geens and \gonzalez shows strong effects (see Fig.~\ref{fig:forestplot}). Nevertheless, all the \textit{effects} agree on its positive trend.

Utilizing model quantization weakly positively affects \textit{inference power draw} of operating DL models. We observe a high conflict between evidence since the reported \textit{effects} span from {indifferent - weakly positive} to positive. This \textit{effect} is the most reinforced after aggregation with a +28\% \textit{belief}.

\subsection{Effects of model quantization with unchanged belief}
\textit{GPU utilization} and \textit{GPU memory utilization} \textit{effects} did not improve their \textit{belief} after the aggregation. Only \gonzalez and \alizadeh reported \textit{GPU utilization}, which accounts for three evidence models. The aggregation shows that model quantization indifferently affects this metric. For \textit{GPU memory utilization}, only two evidence models from \alizadeh reported this \textit{effect}. Their aggregation indicates that model quantization \{positively - strongly positively\} affects \textit{GPU memory utilization}.

\subsection{Effects of model quantization that decreased their belief}
An interesting result of the aggregation is that \textit{inference energy consumption} has decreased belief due to contradictory evidence in the selected studies. Indeed, looking at Fig.~\ref{fig:forestplot}, we observe how this \textit{effect} appears in all the intensity areas between \{IF, WP\} and \{SP\}. Overall, aggregating all this evidence indicates that using model quantization weakly positively affects \textit{inference energy consumption}.

\begin{tcolorbox}
Proposed theory: Model quantization \textit{causes} \textit{positive effects} in DL systems' resource efficiency. \textit{Strongly positive effects} are observed in \textit{storage size} and \textit{GPU energy consumption}. \textit{Inference power draw} is \textit{weakly positively} affected while \textit{\{indiferent - weakly positive\} effects} are observed for \textit{GPU power draw} and \textit{inference latency}. Model quantization also \textit{causes weakly negative effects} on accuracy.
\end{tcolorbox}

\section{Discussion}
\subsection{Effects of model quantization}

A key consideration when applying model quantization is its impact on system correctness. The aggregated evidence suggests that quantization has a \{WN\} \textit{effect} on accuracy. While this might initially appear unfavorable, it is promising, as most systems can handle this minor degradation in accuracy in exchange for the substantial benefits of quantization. Furthermore, we observe how F1 score seems indifferent to model quantization. Nevertheless, more studies reporting this metric are needed to verify this effect since all the aggregated evidence came from a single study. This also shows how different metrics offer distinct perspectives regarding correctness, highlighting the need to select the appropriate metric depending on the system's goal. It also suggests that relying on a single metric to evaluate correctness might miss important system performance aspects.

Despite variations in the study contexts, the aggregated data indicate a consistent trend regarding resource efficiency. Model quantization significantly benefits metrics such as \textit{storage size} and \textit{GPU energy consumption}, independently of the type of DL system or the specific quantization technique employed. A large improvement in \textit{storage size} is expected, given that it is the primary objective of quantization. The \{PO, SP\} \textit{effect} on \textit{GPU energy consumption} appears to result from reduced memory-bound operations---linked to the smaller storage footprint---and fewer overall GPU operations. This interpretation is supported by observed effects in \textit{GPU memory utilization}, \textit{GPU power draw}, and \textit{GPU utilization}.

A relevant finding is the \{IF, WP\} \textit{effect} of quantization on \textit{inference latency}, as it has the second highest \textit{belief} after the aggregation. We want to highlight that, although all the evidence we aggregated reports improvements on \textit{inference latency}, other works not included in the aggregation report otherwise. In this regard, a study by Wallace et al.~\cite{wallaceOptimizationStrategiesEnhancing2025} reported negative effects on \textit{inference latency} when using model quantization. However, the authors remark: ``This time increase is a known factor in the quantization method we use.'' According to the developers of the quantization method, this method is optimized for training rather than inference, explaining its divergence from the broader trend of latency improvements. Another study by Coignion et al.~\cite{coignionGreenMyLLM2024} used the same quantization method with similar latency results. We decided not to include these studies in the aggregation since we consider they do not meet inclusion criteria (iii), given that the method they used is not designed for model inference.

Another relevant outcome is the \textit{effect} of model quantization on \textit{inference energy consumption}. The aggregated \textit{belief} associated with this metric decreased -1.7\% from its peak value, reflecting the heterogeneity of reported intensities. As discussed in Sec.~\ref{sec:results} results span from \{IF, WP\} to \{SP\}. These variations can be attributed to differences in quantization precision. Generally, greater reductions in precision yield stronger improvements in energy efficiency. A separate aggregation with evidence from \sathish and \gonzalez confirms that converting from FP32 to INT8 leads to a strongly positive \textit{effect} on \textit{inference energy consumption}, with an 82\% \textit{belief} and a +12\% increase post-aggregation. Detailed results are available in \cite{EvidenceFactorySynthesisFP32INT8}.

Overall, model quantization weakly compromises correctness while significantly improving resource efficiency. For DL systems operating on embedded devices with strict resource constraints, quantization offers considerable benefits, such as extended battery life, with minimal impact on system accuracy. We identify the quantization precision and quantized components as confounding factors of the aggregation. Based on our observations, smaller quantization precisions lead to lower correctness and higher resource efficiency. The same behavior is observed moving from activations quantization to weights quantization and then to weights and activations quantization. Nevertheless, further research is needed on this topic to verify these hypotheses.

\subsection{Lessons learned on aggregating data strategy studies}

Traditionally, research synthesis in SE has focused on studies involving human subjects. However, an increasing number of studies are now centered on digital entities---such as source code, models, and other software artifacts---as the primary subjects of investigation. All the studies selected for this aggregation fall into this category.

One key observation from reviewing these studies is that \textbf{data strategies often involve multiple independent variables}. In the context of SE for AI, it is common for studies to include: (i) an independent variable of interest (e.g., model quantization), (ii) the DL model acting as the subject under study, and (iii) the dataset on which the model is evaluated. Indeed, all the selected studies followed this pattern. \textbf{This multiplicity of variables complicates the aggregation process, especially since standard synthesis techniques, such as meta-analysis, require studies to follow a common research design}. In contrast, the SSM offers a more flexible framework, accommodating heterogeneity in study design. Nevertheless, the underlying \textit{theoretical structures} must be carefully aligned to ensure compatibility. A key consideration is defining what comprises a single evidence, which influences how data is extracted and mapped into the corresponding evidence models.

A second insight concerns the quality and accessibility of reported data. Only two of the six studies reviewed made their raw data publicly available. \alizadeh provided the data in a readily usable format, and \gonzalez required reviewing the source code. The remaining studies presented only summary statistics (typically means) in tables or visualizations---e.g., \geens reported results exclusively through bar charts. Although aggregation is still possible under such conditions, the reliability of results decreases when the underlying data is incomplete or obscured. Moreover, \textbf{the mere availability of raw data is insufficient. Researchers must ensure it is shared in a clean, well-documented, and accessible format}. This prevents misinterpretation, especially since raw datasets often contain variables not discussed in the publication. These findings underscore the ongoing need for broader adoption of open science practices.

Closely tied to data quality is the overall methodological quality of the studies. In the SSM framework, study quality directly affects the \textit{belief} attributed to the resulting evidence models. Notably, the original SSM quality assessment form was designed with respondent-based studies in mind. When applied to data strategy studies, we found that several questions were not applicable. \textbf{While standard criteria for empirical research still apply, new quality dimensions are needed for these types of studies}. For instance, data quality itself should be assessed, given its central role in study outcomes. Additionally, for studies investigating resource efficiency, the number of repeated runs is an important factor: higher replication yields more reliable results. \textbf{These insights point to a clear need for guidelines specifically tailored to evaluating the quality of data strategy-based research.}

Although there is still work to do, \textbf{data strategy studies offer a unique advantage for evidence aggregation due to their inherently quantitative nature}, often producing structured outputs such as metrics, model performance scores, and correlations. \textbf{This makes them particularly suitable for visual and statistical aggregation techniques, such as forest plots}. Originally developed for summarizing results from randomized controlled trials in medical research, forest plots have since gained prominence in other empirical fields---such as environmental epidemiology---as a powerful tool for synthesizing findings from observational studies. In software engineering, forest plots can be especially valuable for presenting aggregated evidence from multiple data strategy studies, enabling researchers to visualize effect sizes, heterogeneity, and confidence intervals concisely and interpretably. In contrast to human-centered studies, which frequently depend on qualitative data or subjective measures that present standardization challenges, data strategy studies enable more consistent and reproducible aggregation. This characteristic paves the way for scalable and high-precision research syntheses in empirical SE research.

\section{Threats to Validity}

We discuss threats to the validity of our findings and applied mitigations below.

\textit{Internal validity}. Each selected study introduces its internal validity concerns, which we addressed by carefully assessing both the quality and context of each study. One notable challenge lies in the quality assessment process, which was originally designed for respondent-based studies and is not directly applicable to data strategy research. To accommodate this, we modified the assessment instrument by introducing a third response option---``Not applicable''---so that non-relevant questions neither contributed to nor detracted from the base \textit{belief} score.

Another significant threat relates to the presence of different experimental settings within the same study. Several studies evaluated multiple configurations (e.g., quantization method, model architecture, dataset). This intra-study heterogeneity complicates the extraction and interpretation of effects, as improvements may not be directly attributable to the variable under investigation. An example of this design can be seen in \cite{coignionGreenMyLLM2024}, which we did not include in the study, as there is no apparent control over the studied factors. Although SSM is designed to support contextual variability, this diversity must be handled carefully to avoid overgeneralization. To mitigate this, we computed the relative improvement of each configuration independently and only synthesized comparable results. For instance, in \sathish the authors evaluate the impact of using INT8 quantization. The study evaluates three models on a medical image classification dataset and another three models on a medical segmentation dataset. For each of the six models, they measure the inference energy consumption, among other metrics, with and without quantizing the models. To create this study's evidence model, we compared the metrics before and after quantization for each model and computed their relative improvement. For example, we obtain six different values of relative improvement on inference energy consumption. Then, we compute the mean of all six values and the interquartile range. This is then used to establish the intensity of the \textit{effect} on ``inference energy consumption'' and the discount factor. If the study had used multiple quantization methods, we would have repeated the same procedure for each method and created its corresponding evidence model.

\textit{External validity}. To address the threat of omitting relevant primary studies, we systematically searched for empirical studies on model quantization's impact on resource efficiency. A second threat is that we automate the first filtering of papers using an LLM rather than manually inspecting all of them. We have quantified the impact of this risk by manually verifying 50 papers, obtaining a satisfactory recall, and reducing the chance of excluding relevant papers in this step (see~\cite{ReplicationPackage}).

The level at which we perform the aggregation poses another threat since we have observed how the specific quantization method affects the \textit{intensity} of the observed \textit{effects}. However, after extracting all the evidence, we noticed that there was only one quantization method---quantization of weights and activations from FP32 to INT8---reported in more than one evidence model. Thus, we decided to aggregate all the evidence under the ``Model quantization'' construct and provide a general view of the effects of quantization. We acknowledge our findings are not fully generalizable to all quantization methods. However, we expect them to capture the model quantization's main effects and their trends.

\textit{Construct validity}. Interpreting empirical evidence, which can be influenced by researcher bias, also threatens construct validity. To mitigate this threat, one researcher prepared the initial evidence models, which all authors reviewed and validated collaboratively.

To enhance the interpretation of the aggregated evidence, we applied several strategies recommended in prior work~\cite{medeirosdossantosRepresentationAggregationEvidence2013}. For example, since SSM does not inherently account for differences in sample sizes across studies, we refined the \textit{belief} of each \textit{effect} using a discounting mechanism, as described in Section~\ref{sec:evidence-synthesis}. A specific threat here stems from the fact that the discount term used is a novel contribution proposed by the authors. Its formulation aims to reflect the dispersion of reported results relative to their central tendency. While it provides an interpretable way to model variability, its novelty introduces uncertainty regarding its generalizability and reliability. We carefully analyzed its behavior across multiple scenarios to mitigate this and validated its outputs against alternative strategies. Nevertheless, further empirical validation is required to establish its robustness across broader contexts.

The fact that some effects are reported only in a single study (i.e., F1 score and GPU memory utilization) increases the risk of biased measurements. Thus, further evidence of these effects is required to validate our current results.

\textit{Conclusion validity}. As our synthesis relies entirely on what was reported in the primary studies, there remains a residual risk that unreported or missing data could impact our conclusions. We documented the contextual factors of each study to support a nuanced interpretation of our conclusions.

\section{Conclusions and Future Work}
As empirical research in Software Engineering (SE) increasingly adopts data strategy-based studies, effective methods to aggregate such evidence become critical for building theories that can guide future advancements. However, synthesizing findings from these studies presents new challenges, as traditional research synthesis methods have not been fully adapted to accommodate their specific characteristics.

In this study, we applied the Structured Synthesis Method (SSM) to aggregate empirical evidence on the effects of model quantization on deploying Deep Learning (DL) systems. Our synthesis included six primary studies demonstrating that SSM can effectively combine heterogeneous evidence in this emerging area.

The results of our aggregation strengthen the belief in several benefits of model quantization, particularly in improving storage size, inference latency, inference power draw, GPU power draw, and GPU energy consumption. While a weakly negative effect on accuracy was observed, this is considered a manageable trade-off, especially for highly resource-constrained deployment environments. The aggregation also weakened the belief that model quantization affects inference energy consumption due to contradictory results. Understanding this variation is essential. High-belief effects are likely those that have been more extensively and consistently studied, whereas low or negative \textit{belief} values typically stem from limited or conflicting evidence. In addition, evidence on each specific quantization method is scarce, making finer-grained aggregations unfeasible. These areas represent important directions for future research. We encourage the SE community---particularly those working on optimizing DL systems---to further investigate these underexplored effects and quantization methods, contributing to a stronger empirical foundation and more reliable understanding of the impact of model quantization.

We have documented the aggregation process and shared key lessons learned to support future efforts in this area. Although SSM offers flexibility, it remains primarily designed for studies involving human subjects. As part of our future work, we aim to adapt and extend the SSM framework by introducing quality evaluation criteria tailored to data strategy studies. This enhancement will improve the robustness, comparability, and interpretability of future evidence aggregations in SE research. Our proposed extensions to the SSM aim to bridge the gap between emerging empirical methods and synthesis frameworks.

Beyond SSM, future research should explore the adaptation of other synthesis frameworks to better accommodate the distinctive attributes of data strategy studies. As this field progresses, there is a pressing need to systematically characterize these studies. For instance, identifying who or what comprises the ``subjects.'' In the context of data strategy studies, subjects are composed not of human participants, but rather of distinct configurations of data variables, necessitating a reexamination of traditional assumptions regarding the unit of analysis and how this changes the synthesis process. Furthermore, we encourage the formulation of guidelines to evaluate the quality of data strategy studies. These guidelines should consider methodological rigor and the unique challenges of manipulating data-centric artifacts.

\bibliographystyle{IEEEtran}
\bibliography{IEEEabrv,references}

\end{document}